\documentclass[12pt,letterpaper]{JHEP3}
\usepackage[latin1]{inputenc}
\usepackage{bm}
\usepackage{amsmath,amssymb,graphicx,mathrsfs,epsfig}
\usepackage{epstopdf}
\newcommand{\eps}{\epsilon}
\newcommand{\p}{\phi}

\newcommand{\be}{\begin{equation}}
\newcommand{\ee}{\end{equation}}
\newcommand{\bea}{\begin{eqnarray}}
\newcommand{\eea}{\end{eqnarray}}

\newcommand{\kahler}{K\"{a}hler }
\title{Supergravity Higgs Inflation and Shift Symmetry in Electroweak Theory}
\author{Ido Ben-Dayan$^1$ and Martin B. Einhorn$^2$\\
    ${}^1$Department of Physics, Ben-Gurion University,
    P.O. Box 653, Beer-Sheva 84105, Israel
    \\${}^2$Kavli Institute for Theoretical Physics, University of California, Santa Barbara CA 93106-4030, USA\\
    E-mail: idobd@bgu.ac.il, meinhorn@kitp.ucsb.edu}

\abstract{
We present a model of inflation in a supergravity framework in the Einstein frame where the Higgs field of the next to minimal supersymmetric standard model (NMSSM) plays the role of the inflaton. Previous attempts which assumed non-minimal coupling to gravity failed due to a tachyonic instability of the singlet field during inflation. A canonical K\"{a}hler potential with \textit{minimal coupling} to gravity can resolve the tachyonic instability but runs into the $\eta$-problem. We suggest a model which is free of the $\eta$-problem due to an additional coupling in the K\"{a}hler potential which is allowed by the Standard Model gauge group. This induces directions in the potential which we call K-flat. For a certain value of the new coupling in the (N)MSSM, the K\"{a}hler potential is special, because it can be associated with a certain shift symmetry for the Higgs doublets, a generalization of the shift symmetry for singlets in earlier models.  We find that K-flat direction has $H_u^0=-H_d^{0*}.$   This shift symmetry is broken by interactions coming from the superpotential and gauge fields.  
This flat direction fails to produce successful inflation in the MSSM but yields a more interesting model in the NMSSM, even although it does not pass existing cosmological constraints.  We point out that, in building more sophisticated models of this type, one may also need to take into account their implications for axion searches or other elementary particle constraints.
}


\begin{document}

\section{Introduction}

Connecting inflation with low-energy particle physics phenomena is one of the great challenges facing astroparticle physics at present, especially due to the small number of observables and the extremely high energies typically involved in inflationary physics. Embedding inflation in the Standard Model (SM) or its supersymmetric extensions such as the minimal supersymmetric standard model (MSSM) or the next to minimal supersymmetric standard model (NMSSM) is one of the most promising avenues in which to address such a challenge.
The NMSSM was originally introduced to solve the $\mu$-problem of the MSSM. Its main feature is the absence of dimensional coupling constants in the superpotential. (For extensive reviews on the NMSSM and its implications see \cite{Ellwanger:2009dp,Maniatis:2009re}.)

In the SM itself, a viable model of inflation involves non-minimal coupling to gravity \cite{Bezrukov:2007ep}, although such models are beset with all the usual shortcomings of the nonsupersymmetric Standard Model.  Concerning effective field theory, there are still questions concerning the validity of such large non-minimal couplings, as reviewed in \cite{Burgess:2010zq} and references therein. At large fields, there really is no problem with perturbative unitarity.  The main concern, in our opinion, is the possibility of strong gravity in the intermediate region after inflation ends but before the field reaches the electroweak minimum.  Recently, there have been several attempts to generalize the model to the MSSM or the NMSSM within supergravity (SUGRA) where at least some of the theoretical problems were avoided \cite{Einhorn:2009bh,Ferrara:2010yw}. There has also been a recent attempt at using a different non-minimal coupling \cite{Germani:2010gm}. In the MSSM the conclusion was that there is no viable model for the Higgs field as the inflaton even with non-minimal coupling, while in the NMSSM such a scenario is possible. It was shown that the original model in the NMSSM  \cite{Einhorn:2009bh} failed due to a tachyonic instability in the singlet field $S$ \cite{Ferrara:2010yw}. Several solutions to the instability were suggested \cite{Lee:2010hj,Nakayama:2010sk}.\footnote{For a review of the effects of non-renormalizable terms and how they may
affect instabilities in the potential, see \cite{Mazumdar:2010sa}.}
Alternatively, there have been attempts to construct other inflation models within the (N)MSSM with or without non-minimal coupling to gravity as well as within the SUGRA framework or outside it.  In most of them, the inflaton was some additional fundamental field which is a singlet of the SM gauge group or some composite field, for example \cite{Allahverdi:2006cx, Allahverdi:2007wt}. For a comprehensive review of such models, see \cite{Mazumdar:2010sa}.

Within the SUGRA framework, there is the well-known large-$\eta$ problem. Due to the functional structure of the potential, it tends to be very steep and block inflation. One common solution is to resort to a "small field model", which means that the slow-roll conditions are met in a small region in field space, $\Delta\phi\lesssim 1$. For example, inflation starts near some extremum of the potential. As was demonstrated in \cite{BenDayan:2008dv,BenDayan:2009kv,German:1999gi,German:2001tz}, these models avoid the $\eta$-problem and have a variety of observable consequences including detectable gravitational waves and detectable spectral index running.

Another common solution to the $\eta$-problem is to invoke a shift symmetry which protects the inflaton direction from steepening \cite{Kawasaki:2000yn,Kasuya:2003iv}, and get a large field model $\Delta \phi \gg 1$. As explained in \cite{Baumann:2010ys}, such a symmetry may be broken at the Planck-scale and receive dangerous corrections which could spoil inflation. The main danger comes from the fact that the symmetry is a global or discrete symmetry and, according to lore, cannot be a true symmetry at the Planck scale. The authors in \cite{Baumann:2010ys} suggested a way to desensitize inflation from the dangerous corrections at the Planck scale by coupling it to a conformal sector.

The model we present here uses the SM gauge group but adds an additional term in the K\"{a}hler potential.  The model assumes minimal coupling to gravity and uses canonical kinetic terms, so it avoids the issues of the validity of the effective field theory . For the model to have any chance of working, the coupling constant of the additional term $\zeta$ must be close to unity, raising the question of naturalness. We generalize the shift symmetry in a manner consistent with the global $SU(2)\otimes U(1)$ and making the value $\zeta=1$ technically natural.  Like the earlier shift symmetry for singlets, this shift is broken by terms in the superpotential, as well as by gauge interactions.  The inflaton is the Higgs field, so detection of the Higgs and/or SUSY in the LHC will be another constraint on the model. An interesting part of the model is in the Peccei-Quinn approximation. In this limit there is a similarity between the constraints coming from inflation and the constraints coming from axion searches.   
Unlike models involving only singlet inflatons or hidden sector fields, this illustrates that models of this type will have to pass much more stringent tests than just cosmological observations.

The net result is a chaotic inflation model with a self-interacting $\phi^4$ scalar field.
This is 
essentially ruled out
by CMB observations \cite{Komatsu:2010fb,Dunkley:2010ge}.
However, there are interesting ingredients of the construction, such as a new shift symmetry and a different form of \kahler potential that may provide useful building blocks for future models. The failure of such simple Higgs inflation models to explain the CMB is one of the reasons to consider more complicated scenarios such as non-minimal coupling to gravity or non-canonical kinetic terms.

We use the usual slow-roll approximation for a single real field, except in places where we would use a more rigorous SUGRA treatment. Working in Planck units, where the reduced Planck mass $M_p=1$ this means the slow-roll parameters  and number of e-folds are:
\be
\epsilon=\frac{1}{2}\left(\frac{V'}{V}\right)^2, \quad \eta=\frac{V''}{V}, \quad \xi=\frac{V'''V'}{V^2}, \quad N(\phi) = \int_{\p_{END}}^{\phi}\frac{d\p}{\sqrt{2 \eps(\p)}},
\ee
where prime denotes differentiation with respect to the inflaton. Inflation ends when $\epsilon \simeq 1$.
The cosmic microwave background (CMB) observables evaluated as cosmological scales leave the horizons are:
\be
n_s=1+2\eta-6\epsilon, \quad r=16 \epsilon, \quad \frac{d n_s}{d \ln k}=16\epsilon \eta-24\epsilon^2-2\xi.
\ee

As was attempted in \cite{Einhorn:2009bh,Ferrara:2010yw}, we concentrate on achieving a 
model for inflation from the Higgs sector.
Since we will deal with inflation near the Planck scale, we suppress the soft SUSY breaking couplings and as in \cite{Einhorn:2009bh,Ferrara:2010yw} assume that after inflation, the system will roll down to the SUSY breaking vacuum of the (N)MSSM in accord with low energy phenomenology.

The paper is organized as follows: we begin with a short review of why the original NMSSM non-minimal coupling to gravity model failed. This is a derivation stemming from a general K\"{a}hler geometry approach as was demonstrated by \cite{BenDayan:2008dv,Badziak:2008yg,Covi:2008cn}. We then present the additional new ingredient to the K\"{a}hler potential in section $3$. In section $4$ we show that this can be associated with a new shift symmetry of the K\"{a}hler potential, that is compatible with the global $SU(2)\otimes U(1)$ symmetries, forcing the coupling constant to be unity.

In section $5,$ as an exercise we apply this idea to the MSSM and show that it still does not give rise to a viable model of inflation.  In section $6,$ we embed the model in the NMSSM and find 
an inflationary model that can be compared with CMB observables. We discuss the qualifications of the model in section $7$ and conclude in section $8$.  Throughout the paper we consider just $N=1$, $d=4$ Einstein-frame SUGRA. During work on this manuscript, we found out that the idea of adding a holomorphic part to the K\"{a}hler potential and addressing some of its phenomenological consequences was presented independently in \cite{Ferrara:2010in,Kallosh:2010ug}.

\section{NMSSM Model and the Tachyonic Instability}

We can examine the NMSSM model in the form originally considered in \cite{Einhorn:2009bh} by considering the Einstein frame and by analyzing the model according to the condition derived in \cite{Covi:2008cn}. According to the theorem, a necessary condition for a successful model of inflation can be phrased in terms of the sectional holomorphic curvature as follows.
As usual \cite{Wess:1992cp}, define the function,
\be
G(\phi_i,\bar{\phi}_i)=K(\phi_i,\bar{\phi}_i)+\ln|W(\phi_i)|^2,
\ee
where $K$ is the usual K\"{a}hler potential and $W$ is the superpotential, in terms of which the F-term of the potential becomes
\be
V_F=e^G[K^{i\bar{j}}G_i G_{\bar{j}}-3],
\ee
where $G_i={\partial G}/{\partial\phi^i},$ and $K^{i\bar{j}}$ is the inverse of the K\"{a}hler metric.
Define a unit vector along the goldstino direction by:
\be
f_i=\frac{G_i}{\sqrt{G_j G^j}}.
\ee
The sectional holomorphic curvature along this direction is then given by:
\be
R(f^{i})\equiv R_{i\bar{j}k\bar{l}}f^i f^{\bar{j}} f^k f^{\bar{l}}
\ee
The necessary condition for inflation is then:
\be \label{eq:constraint}
R(f^i)\lesssim \frac{2}{3}\frac{1}{1+\gamma}; \quad \gamma=\left(\frac{H}{m_{3/2}}\right)^2
\ee
The theorem deals only with the $F$-term  potential and of course $D$-terms may change it. However, in the model under discussion, the nearly flat direction has a vanishing $D$-term \cite{Einhorn:2009bh,Ferrara:2010yw}, so the theorem should hold.

The NMSSM model has K\"{a}hler potential $K$ and superpotential $W$ given by
\bea
K&=&-3\ln[ 1 +  (\chi H_d H_u +h.c.)/2-(|H_u|^2+|H_d|^2+|S|^2)/3)]\\
W&=&\lambda S  H_d H_u+\rho S^3/3.
\eea
As explained later, one may take the three couplings $\chi, \lambda,$ and $\rho$ all positive.
The neutral parts of the Higgs doublets are parameterized in the standard way:
\be \label{eq:vevs}
H^0_d=h \sin \beta \exp{i\alpha_d},~H^0_u=h \cos \beta \exp{i\alpha_u},
\ee
where, by definition of the phase factors, $0\le\beta\le\pi/2.$

The vanishing of the $D$-term corresponds to $\beta=\pi/4.$  One may then show that, for $S=0,$ the potential has a local minimum in $\alpha\equiv\alpha_u+\alpha_d$ for $\alpha=\pi.$  Thus, the candidate inflaton trajectory takes place along $h$ at $S=0,\beta=\pi/4$ with $h \ll 1 \ll \chi h$. Along this direction the superpotential vanishes, which makes the calculations a bit easier.   Direct calculation show that
\be
R(f)=\frac{2}{3}+\frac{2}{9\chi}+\frac{4}{3\chi^2 h^2}+\mathcal{O}\left(\frac{1}{(\chi h)^3}\right)
\ee
Hence, $R(f)>2/3,$ so the necessary condition is not fulfilled, and the model has a direction that is too steep to support inflation. This is associated with the tachyonic instability found by \cite{Ferrara:2010yw}.

\section{Canonical K\"{a}hler Models}

When the K\"{a}hler potential is of the form $K=\sum \phi_i\bar{\phi}_i+F(\phi_i)+\overline{F}(\bar{\phi}_{\bar{i}})$, field space is flat, i.e. $R(f)=0,$ and the necessary condition eq.~(\ref{eq:constraint}) is easily met.
\footnote{Using
K\"{a}hler invariance, the holomorphic terms could be incorporated into the superpotential, but, for our purposes, it will be more convenient to retain this more general form.}  In order to have large field models in SUGRA, the
 inflaton is usually protected by a shift symmetry to avoid the ubiquitous $\eta$-problem.
For example, the canonical K\"{a}hler potential of a single chiral superfield $\Phi$ in SUGRA is
\be
K_1=\Phi\bar{\Phi},
 \ee
 so for large values of $\Phi$ one gets a steep potential  $V=e^{K}[\cdots]$ because of the exponential. However, the kinetic terms of the chiral super-fields are $K_{i\bar{j}}\partial\Phi^i\partial\bar{\Phi}^{\bar{j}}$. Thus, the  K\"{a}hler metric $K_{i\bar{j}}$ is independent of the choice of holomorphic function $F(\phi_i).$
If, in the simple single field case, one chooses $F(\Phi)=\Phi^2/2$, the K\"{a}hler becomes:
\be
K_2=\frac{(\Phi+\bar{\Phi})^2}{2}
\ee
The kinetic terms are identical to those of $K_1;$ however, now the K\"{a}hler potential possesses a shift symmetry $\Im{\Phi}\rightarrow \Im{\Phi}+\Im{a},$ where $a$ is some complex number.  Since $K$ is independent of the imaginary part of $\Phi,$ it becomes a natural candidate for the inflaton.  With a suitable choice of the superpotential, inflation is possible\footnote{To be exact in \cite{Kawasaki:2000yn} $K=\frac{1}{2}(\Phi+\bar{\Phi})^2+X\bar{X}$ and $W=m X\Phi$.} \cite{Kawasaki:2000yn}.

Turning to the Higgs sector of the (N)MSSM, we begin with
\bea \label{eq:kahler}
K_{MSSM}=|H_u|^2+|H_d|^2+\zeta H_dH_u+h.c.;\\
K_{NMSSM}=|H_u|^2+|H_d|^2+|S|^2+\zeta H_d H_u+h.c.
\eea
Despite similarities in appearance with earlier nonminimal supergravity constructions, we are assuming {\it minimal coupling} to gravity.     Suppressing the charged fields, and supposing that we seek inflationary solutions that have classical backgrounds with large field values,
 it is clear that neither $H_u^0=0$ nor $H_d^0=0$ are candidates for such directions, so we may parameterize them as in eq.~(\ref{eq:vevs}), giving
\bea \label{eq:kahler2}
K_{MSSM}&=&h^2(1+\zeta\sin{2\beta}\cos{\alpha});\\
K_{NMSSM}&=&h^2(1+\zeta\sin{2\beta}\cos{\alpha})+|S|^2,
\eea
where $\alpha\equiv\alpha_u+\alpha_d.$   For the time being, we will ignore the singlet field $S.$
In this parametrization, the K\"{a}hler potential is expressed as a function of three real fields  $h, \beta, \alpha.$
The gradient of $K$ points in the direction of greatest increase of $K$ (and is normal to a constant $K$-surface.)
\be
\Big(\frac{\partial K}{\partial h},\frac{~~\partial K}{h\,\partial\beta},\frac{~~\partial K}{h\,\partial\alpha}\Big)=
h\Big(2(1+\zeta\sin{2\beta}\cos{\alpha}),2\zeta\cos2\beta\cos\alpha,-\zeta\sin2\beta\sin\alpha\Big)
\ee
At an extremum, the gradient will vanish and, generically, that only occurs at isolated points.  For inflation, we seek a ``degenerate" circumstance in which the gradient vanishes on at least a one-parameter curve,\footnote{One physically may be able to accept a sufficiently slowly varying such curve, and we will discuss this possibility later.} and, in the absence of some symmetry, we expect it would occur only under special situations or for particular values of the parameters of the K\"ahler potential.   We call such a curve a ``K-flat direction."    It is clear that  $\sin{2\beta}\cos{\alpha}$ must not vanish for the first component to be small for a wide range of $h.$  Looking at the other two components, the only circumstance where they vanish while satisfying this constraint is for $\beta=\pi/4, \alpha=\pi.$  Then the first component is $2h(1-\zeta),$ which vanishes only if $\zeta=1.$  Thus, this scenario is strictly possible only in a neighborhood  of a particular value of the coupling $\zeta.$  (This would appear to be rather fine-tuning, but we shall see in the next section that this value corresponds to a shift symmetry of $K.$)

For the NMSSM, we must include the field $S$ as well, but one can see that $S=0$ is stationary and stable.  Further, there are no other curves for large $|S|$ and $h$ which satisfy these necessary conditions.

Therefore, we have found a direction $\beta=\pi/4,\alpha=\pi$ along which K is stationary for a particular value of the coupling $\zeta$.     For this to be a stable trajectory, it must be a minimum, and it is easy to see that this is true.   Further, for $\zeta=1,$ all higher derivatives in $h$ vanish as well, so the field $h$ would seem to be a candidate for an inflaton, provided the full potential is positive there.  It must also be that superpotential has a form that does not cause to the full potential to vary much along this direction, and that too is a strong constraint on models.

We shall discuss these possibilities in more detail in the following sections, but we would like to make a couple of observations beforehand.  First, in general, $\zeta$ is a running coupling that depends on a normalization scale, $\zeta=\zeta(t).$  Therefore, choosing a particular value of $\zeta$ corresponds at best to the value at a particular normalization scale.  Unless one can find a symmetry that requires $\zeta=1$ in eq.~(\ref{eq:kahler}), there is nothing special or natural about this particular $\zeta.$  This provides motivation to generalize the shift symmetry of the singlet model, which we will do in the next section. Second, one could add terms involving the singlet field $S,$ so that $K_{NMSSM}$ is a function of $(S+\bar{S})^2$ only, just as in the single field inflation model above.  So it would be easy to incorporate such a construction into the NMSSM and use the singlet $S$ as the inflaton.  It would remain to be investigated whether the singlet $S$ could be identified simultaneously with the inflaton and, at the same time, the field whose VeV determines the Higgs mass to be of electroweak size.  In this paper, we are more interested in the possibility of having the inflaton identified with the Higgs fields, and we will not pursue this course here.

\section{Shift Symmetry for Doublets}

The (N)MSSM fields are special inasmuch as they allow a generalization of shift symmetries from singlet fields to non-singlet fields in a manner consistent with global $SU(2)\otimes U(1).$  Recall that $H_u$ and $H_d$ are both doublets but have opposite hypercharges.  Therefore, $H^*_d$ has the same hypercharge as $H_u.$   It transforms as the conjugate representation $\underline{2^*}$ under $SU(2)$, but, as is well-known, $i\sigma_2 H^*_d$ transforms as a doublet $\underline{2}$ (where $\sigma_2$ is the Pauli matrix.)
Consider the shift defined by
\bea \label{eq:shift}
H_u\rightarrow H_u+C,& \quad&H_d\rightarrow H_d - i \sigma_2 C^*\\
S&\rightarrow& S\nonumber
\eea
where $C$ is a constant $SU(2)$ doublet to which we assign the hypercharge of $H_u.$   This symmetry takes advantage of two facts: (1) the spinor representations of SU(2) are pseudoreal, so that $\underline{2^*}\cong\underline{2},$  and (2) a constant field is the unique superfield that is both chiral and antichiral~\cite{Wess:1992cp}, so that this shift is compatible with supersymmetry\footnote{Such a
shift symmetry can obviously be defined for any field $\phi$ whenever $\phi*$ is a group representation equivalent to that of $\phi.$  In particular, it can always be applied to a chiral field in the adjoint.}.  The first property is the reason that one can get by with a single Higgs doublet in the SM, while the second is one reason for the need for two Higgs doublets in supersymmetric models.  Thus, this shift symmetry is compatible with a global $SU(2)\otimes U(1)$ symmetry.   Given this shift symmetry, one may check that $H\equiv H_u - i\sigma_2 H_d{^*}$ is invariant\footnote{Unlike the single field case, the shift symmetry requiring
$H\equiv H_u - e^{i\gamma}  i\sigma_2 H_d{^*}$ invariant, for any phase angle $\gamma,$ is not distinct.   One may simply absorb the phase into $H_d$. However, one cannot require invariance simultaneously under both eq.~(\ref{eq:shift}) and $H_u\rightarrow H_u+C,~H_d\rightarrow H_d + i \sigma_2 C^*.$}.  Note that this applies to the full doublet, so that the invariance applies to both charged and neutral modes.

If we impose this shift symmetry on the K\"{a}hler potential, then it  must be constructed from $H\equiv H_u - i\sigma_2 H_d{^*}.$   One can show that
\be
H^\dagger H=H_d^\dagger H_d + H_u^\dagger H_u+(H_dH_u+h.c.).
\ee
This is precisely the Higgs dependence of the  K\"{a}hler potentials in eq.~(\ref{eq:kahler}) for $\zeta=1.$   Focusing on the neutral fields, this means that the K\"{a}hler potential is a function of the linear combination $H_u^0 + H_d^{0*}.$   In terms of our parametrization, $\zeta=1$ can be K-flat only for $\beta=\pi/4, \alpha=\pi,$   corresponding to $\langle H\rangle\!=\!0.$  This is the {\it converse} of what we showed previously, where these values emerged as the stable minima defining the inflaton direction.  Therefore, it seems that the classical fields will dynamically relax toward the point having this shift symmetry. This is rather remarkable, since the shift symmetry is broken by the gauge interactions and by terms in the superpotential involving these fields.

The shift symmetry has additional physical consequences because it means that the flat direction (both charged and neutral) will get masses only because of the breaking of the shift symmetry by gauge and other interactions.

\section{MSSM}

Mostly by way of a preliminary exercise, let us consider the MSSM with a particular choice of superpotential.
\bea
K_{MSSM}=|H_u^0|^2+|H_d^0|^2+\zeta H_d^0H_u^0+h.c., && W_{MSSM}=\Lambda+\mu H_d^0H_u^0\\
V_F=e^K[K^{i\bar{j}}D_{i}W D_{\bar{j}}\bar{W}-3|W|^2],&&V_D=\frac{g_1^2+g_2^2}{8}(|H_u^0|^2-|H_d^0|^2)^2,\\
V&=&V_F+V_D
\eea
where $i,\bar{j}$ run over both chiral superfields, and we have assumed we may ignore the charged fields $H_u^+=H_d^-=0$.
Note that the K\"{a}hler metric is trivial, $K_{i\bar{j}}=\delta_{i\bar{j}}.$
Without loss of generality, we may take  $\zeta>0$ and $\mu>0.$  The constant
$\Lambda$ may be complex, but in fact, one can verify that it does not affect our conclusions so, for simplicity, we will set $\Lambda$ to zero.

Parameterizing $H_u^0$ and $H_d^0$ as in eq.~(\ref{eq:vevs}), the K\"{a}hler potential takes the form given above in eq.~(\ref{eq:kahler2}).    The full potential then takes the value
\bea
V_F&=&\frac{\mu^2h^2}{4} e^{h^2\omega} \Big[4+h^2(4\omega-4+\sin^22\beta)+h^4\sin^22\beta\big( \zeta^2-1+2\omega\big)\Big]  \\
V_D&=&\frac{g_1^2+g_2^2}{8}h^4\cos^22\beta,
\eea
where we defined $\omega\equiv 1+\zeta\sin2\beta\cos\alpha$, so that $K=h^2\omega.$
We have already determined the candidate K-flat trajectory exists only for $\beta=\pi/4, \alpha=\pi,$ and we will now see that, fortunately, these values also correspond to stable minima of the full potential. We also expect to have $\zeta\approx1.$
Note that the only dependence on $\alpha$ is through $\omega$ and that the full potential $V\equiv V_F+V_D$ can be regarded as a function of $(b,\omega),$ where $b\equiv\sin2\beta$.  Therefore,
\bea \label{eq:extremealpha}
\frac{\partial V}{\partial\alpha}&=&-\zeta \sin2\beta\sin\alpha \frac{\partial V_F}{\partial\omega}\\
 \label{eq:extremebeta}
\frac{\partial V}{\partial\beta}&=&2\cos2\beta\Big[\frac{\partial V}{\partial b}+\zeta\cos\alpha \frac{\partial V_F}{\partial\omega}\Big].
\eea
If $V$ is to be stationary for all values of $h$, these expressions suggest (and one can verify) that $\beta=\pi/4$ and $\alpha=\pi$ is the only possibility in the K-flat direction.  Note that at the extremum, $V_D=0.$ Inputting these values, the formulas become simpler, with
\be
V=\frac{\mu^2h^2}{4} e^{h^2\omega} \Big[(2+\omega h^2)^2-3h^2\Big]
\ee
with $\omega=1-\zeta.$    For coupling $\zeta=1,$ the terms in brackets reduce to $[4-3h^2],$ which is clearly neither slowly varying nor positive for large $h^2.$  For $\omega\ne0,$ the potential is bounded from below, but its minimum is at a very large $\omega h^2,$ for small $\omega.$   In order to avoid having the $\eta$-parameter unacceptably large, the exponential requires $h^2\omega<1$ for large $h,$ so the range of $\zeta$ is quite restrictive.  Not surprisingly, this remains unacceptable as a model for inflation when $\omega\ne0.$  Last but not least, one might consider a small field model of the new-inflation type. 
For this, one needs to reintroduce a constant term $\Lambda$ into the superpotential to account for the scale of the vacuum energy.  Small field models of this type have been attempted, but they all fail because $\mu$ must be much larger than the electroweak scale to account for inflation.

\section{NMSSM}

Now consider the same sort of approach in the framework of the NMSSM where, we shall see, one can get an acceptable model.
\bea
K&=&|H_u|^2+|H_d|^2+|S|^2+ (\zeta H_dH_u+h.c.)\\
W&=&\lambda S  H_d H_u+\frac{\rho}{3} S^3.
\eea
In general, the coupling constants $\zeta,$ $\lambda,$ and $\rho$ are all complex, but let us determine which of their phases are observable.   Since the overall phase of the superpotential $W$ is unobservable, we may assume $\lambda>0.$  Next, let us recall that two quantum field theories are equivalent if any complex field $\Phi$ is replaced by $\exp(i\theta)\Phi$ for an arbitrary constant phase factor $\theta.$  The \kahler potential $K$ depends on the Higgs fields through the product $H_dH_u,$ so if we multiply each superfield by a phase, it is equivalent to replacing $\zeta$ by $\zeta\exp(i\gamma),$ where $\gamma=\gamma_d+\gamma_u$ is the sum of these phases.  Therefore, we may choose $\zeta>0.$    This leaves only $\rho$ possibly complex.  If we replace the field $S$ by $S\exp{i\theta}$, then the superpotential takes the form
\be
W=e^{i\theta}\big(\lambda S H_d H_u+\frac{\rho}{3} e^{2i\theta} S^3 \big).
\ee
Again, the overall phase of $W$ is unobservable, and the phase of $\rho$ may be absorbed by choice of $\theta,$ so we may assume
$\rho>0$ as well.  Thus, without loss of generality, all three coupling constants may be taken to be real and positive.\footnote{Note that no assumptions about CP-invariance nor restrictions on the phases of the fields have been made.}

Clearly, the K-flat directions will be the same as in the MSSM, except that we must also demand $S=0.$ The D-term $V_D$ is the same as before, but the superpotential and therefore the F-term $V_F$ is different.  It is easy to see that the first corrections about $S=0$ are quadratic in $S$, so that $S=0$ will be an extremum.  We shall first analyze the potential for $S=0$ and then discuss the behavior in $S$.  The value of the potential at $S=0$ is
\be \label{eq:vszero}
V=\frac{h^4}{4}e^{h^2\omega}\lambda^2 \sin^2 2\beta+\frac{g^2}{4}h^4\cos^2 2\beta,
\ee
where $g^2\equiv(g_1^2+g_2^2)/2.$
As in the MSSM, we may regard $V$ as a function of $(b,\omega)$, so we seek extrema as in eqs.~(\ref{eq:extremealpha}) and (\ref{eq:extremebeta}).  As before, these equations suggest that the only extremum independent of $h$ in the K-flat direction has
   $\beta=\pi/4$  and $\alpha=\pi.$  Thus, once again we find this scenario is possible only for coupling $\zeta=1,$ so we find for $S=0,$ the value of the potential is
\be \label{eq:cw}
V=\frac{\lambda^2 h^4}{4}>0; \quad H_V=\frac{\lambda h^2}{2\sqrt{3}}.
\ee
So we have a candidate model of chaotic inflation with a quartic potential.  For future reference, we have given the Hubble scale $H_V$  associated with this value of $V.$

However, we still must determine whether this  extremum is stable against fluctuations.   For this purpose, we need the second-derivatives at the extremum.  From eqs.~(\ref{eq:extremealpha}) and (\ref{eq:extremebeta}), we find
\bea \label{eq:curvalpha}
\frac{\partial^2 V}{\partial\alpha^2}&=&-\sin2\beta\cos\alpha \frac{\partial V_F}{\partial\omega}=+\frac{\partial V_F}{\partial\omega}\\
 \label{eq:curvbeta}
\frac{\partial^2 V}{\partial\beta^2}&=&-\sin2\beta \Big[\frac{\partial V}{\partial b}+\zeta\cos\alpha \frac{\partial V_F}{\partial\omega}\Big]=
\Big[\frac{\partial V_F}{\partial\omega}-\frac{\partial V}{\partial b}\Big],
\eea
where in the last step on each line, we evaluated the expressions at the extremum.  One easily sees that the off-diagonal, mixed-derivative term vanishes at the extremum, so for stability, we need each of these expressions to be positive.  By explicit evaluation, we find
\be
\frac{\partial^2 V}{\partial\alpha^2}=\frac{\lambda^2 h^6}{4}>0, \quad \frac{\partial^2 V}{\partial\beta^2}=\frac{\lambda^2h^6}{4}+\frac{(g^2-\lambda^2)h^4}{2}>0
\ee
for all $h\ne0$ provided $g^2>\lambda^2.$ The masses associated with each of these fluctuations turn out to be
 \be
 m_A^2 =  \frac{\lambda^2 h^4}{2},  \quad m_B^2=\frac{\lambda^2h^4}{8}+\frac{(g^2-\lambda^2)h^2}{4}.
 \ee
The constraint, $g^2>\lambda^2,$ on the couplings is easily satisfied, since the gauge couplings are $O(1)$,  whereas, as we shall see, $\lambda$ will have to be very small to satisfy cosmological constraints.  Before discussing them, we must first evaluate the curvature in $S.$   Evaluating the potential to $O(S^2),$ we find
\be
V = \frac{\lambda^2 h^4}{4}+\lambda h^2\big(\lambda |S|^2-\rho\, \Re\{S^2\}\big).
\ee
Because this is second-order in $S,$ we have set the other fields equal to their extremal values.  To separate the independent modes, we write $S\equiv(s_1+ i s_2)/\sqrt{2}$, so that the terms in $S$ become
\be
V = \frac{\lambda h^2}{2}\big[(\lambda-\rho)s_1^2+(\lambda+\rho)s_2^2\big]\equiv \frac{m_1^2}{2}s_1^2+\frac{m_2^2}{2}s_2^2.
\ee
 provided $\lambda>\rho.$
These values of the masses are dangerously small.\footnote{We thank A. Linde for pointing this out.}  For these fluctuations to dampen within a Hubble time, we must have these masses $m_i^2> H_V^2,$ or $1\pm\rho/\lambda>h^2/12,$ which is certainly not the case for  $h$ sufficiently large.

To sum up the preceding results, in the special case where the coupling constant $\zeta=1,$ it appears that the K-flat direction is stable against fluctuations in the neutral fields  provided $g>\lambda>\rho>0.$  The potential along this direction is simply given by eq.~(\ref{eq:cw}).  However, the scalar modes may be too small compared to the Hubble scale $H_V,$ giving large fluctuations and, perhaps, large nonadiabatic and non-Gaussian effects.  In an Appendix, we show that the fluctuations due to charged Higgs fields do not alter these conclusions, but, not too surprisingly, the charged fluctuations along the flat direction are also worrisomely small.

Models of this type with a quartic potential are well known. (For reviews, see for example \cite{Shafi:2007jn, Baumann:2009ds}.)
For a model satisfying the COBE normalization, one must have $\lambda \simeq 10^{-7}
- 10^{-6}.$
The predictions for $N$ e-folds of inflation are:
\be
n_s\simeq
1-\frac{3}{N}|_{N=60}=0.95,~~r\simeq\frac{16}{N}|_{N=60}\simeq0.26,~~\frac{d n_s}{d \ln k}\simeq-\frac{3}{N^2}|_{N=60}\simeq-8\times10^{-4}.
\ee
These cosmological parameters are already tightly constrained by CMB observations.  The WMAP collaboration five-year analysis \cite{Komatsu:2008hk} showed that, assuming negligible running and $n_s\approx 0.95$ gives a tight bound on $r$ that rules out the $\phi^4$ model by more than $99\%$ confidence level (CL). The WMAP seven-year analysis \cite{Komatsu:2010fb} as well as the recent Atacama cosmology telescope measurements (ACT) \cite{Dunkley:2010ge}, did not change the situation. In brief, it seems the model in its current version is
strongly disfavored, and improved data from the PLANCK satellite observations, already underway, seem destined to remove any possible lingering doubts.

We have given their numerical values for $N=60,$ which corresponds to a range of the inflaton $h$ from about $h_i\gtrsim 22$ initially down to $h_f\approx \sqrt{8}$ at the end of inflation (where $\epsilon\approx1.$)   For this range, the scalar modes $m_i^2$ are small compared the Hubble scale $H_V$ at the start of inflation and comparable to $H_V$ at the end.  It remains to be determined whether these are phenomenologically acceptable values, although it is possible to increase them by adding an $(S^\dagger S)^2$ term to the \kahler\ potential \cite{Lee:2010hj, Kallosh:2010ug}.
The potential as a function of $h$ and $s_1=\Re\{S\}$ with $(s_2=0,\beta=\pi/4,\zeta=1,\alpha=\pi,\rho=0,\lambda=10^{-7})$ is presented in figure $1$.
\FIGURE[t]{
\scalebox{1.3}{\includegraphics{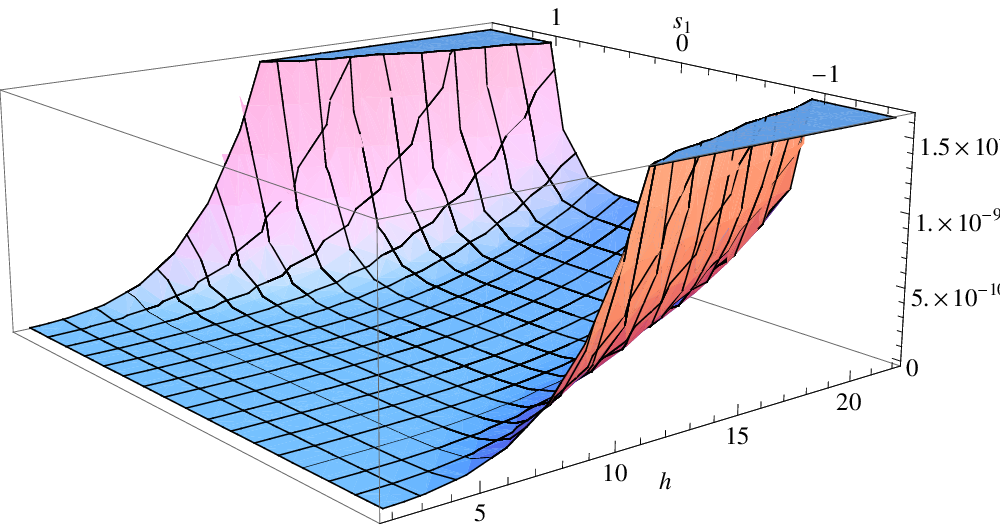}}
\caption{The potential $V(h,s_1)$ in the inflationary region with the following values of parameters and fields: $(s_2=0,\beta=\pi/4,\zeta=1,\alpha=\pi,\rho=0,\lambda=10^{-7})$. }\label{figure1}
}
Assuming $\lambda \gtrsim \rho,$ then the model must be subjected to LHC experiments and CMB observations.

The model is especially interesting in the Peccei-Quinn limit. Since the trilinear term $\rho S^3$ does not play a role in the inflationary dynamics, consider the situation where $\rho \ll \lambda$.  In the global susy limit, there is an approximate Peccei-Quinn $U(1)$ symmetry, so the theory should have an axion. This scenario was advocated in \cite{Feldstein:2004xi} as a solution to the strong CP problem and the $\mu$-problem. As stated in \cite{Maniatis:2009re}, current bounds from axion searches yield $10^{-10}<\lambda<10^{-7}.$   Coincidentally, the value of $\lambda$ needed for inflation is in the same region of the parameter space where the axion is expected.
 The idea is the following, the $\zeta$ term in the \kahler potential breaks the Peccei-Quinn or $\mathbb{Z}_3$ symmetry which the NMSSM originally possesses. Hence there is no danger of domain-wall problem\cite{Ferrara:2010yw}. After inflation is over and the universe settles into the electroweak minimum gravity is extremely weak and we are left with global SUSY NMSSM. If the values of the coupling constants did not change too much during this evolution, the theory should have an axion due to the approximate PQ symmetry and the model is subjected to axion searches experiments. To summarize, for this model to work it should pass three different experimental setups: CMB observations, LHC experiments, and axion searches. If it passes all three tests, it is a striking advantage of the model compared to other models which just meet CMB requirements. Conversely, the model can be easily ruled out.

 \section{Qualifications and Radiative Corrections}

Although the choice of coupling $\zeta=1$ can be motivated by a shift symmetry that is compatible with the global
$SU(2)\otimes U(1)$ structure of the electroweak theory, it is broken by gauge interactions and by the terms in the superpotential involving these fields.  How close must $\zeta$ be to $1$ in order to have a 
chance at an acceptable model?
Returning to eq.~(\ref{eq:vszero}), setting $\alpha$ and $\beta$ at their values at the minimum, the potential is simply
\be
V=\frac{h^4}{4}e^{h^2\omega}\lambda^2,
\ee
where $\omega=1-\zeta.$ One may recalculate the cosmological parameters with $\zeta\ne1.$  For example,
\bea
\sqrt{2\epsilon}=\frac{V'}{V}=\frac{4}{h}+2h\omega\\
\eta=\frac{12}{h^2}+18\omega+4\omega^2 h^2
\eea
From the mild requirements $N=60$, $1 \geq n_s \geq 0.9$ and $r\leq 0.54$, one finds that $-1\% <\omega<0.2\%$ and that $h$ varies during inflation by about a factor of $10.$

The preceding discussion has been classical, and there is much to say and to learn about quantum corrections.  There are two kinds to worry about:  First, $\zeta$ will be scale dependent $\zeta=\zeta(t),$ and second, there will be corrections to the effective potential (and more generally, to the effective action.)   To understand how $\zeta$ runs, it is easier to do a K\"ahler transformation to include it in the superpotential.  It will enter in nonpolynomial form  $\exp(\zeta H_d H_u) W.$   Nevertheless, by the nonrenormalization theorem, it will not be renormalized.  Therefore, the running of $ \zeta$ is determined by the running of the fields, {\it viz.,\,}
\be
\frac{\partial \zeta}{\partial t}\equiv\zeta\beta_{du}(t),~~ \beta_{du}(t)\equiv \gamma_d(t)+\gamma_u(t),
\ee
where $\gamma_d$ and $\gamma_u$ are the anomalous dimensions of the fields $H_d$ and $H_u,$ respectively.
As an aside, if we added a term $\mu H_d H_u$ to the superpotential, then the same is true for $\mu$, {\it viz.,}
$\beta_\mu=\mu\beta_{du}.$\footnote{In fact, if one added a nonminimal coupling $\chi R H_d H_u$ to the theory, then $\beta_\chi=\chi\beta_{du},$ in the approximation that the background metric can be treated classically, i.e, that feedback on the background metric can be ignored.} In any mass-independent renormalization scheme, such as $\overline{DR},$ $\beta_{du}$ is a function of the dimensionless coupling constants other than $\zeta$ in this model, the largest of which will be the gauge couplings and, when extended to include other fields, the top-quark Yukawa coupling.

We have seen that $\omega=1-\zeta$ is tightly constrained by the cosmological parameters.  In our illustration above, we found the range of $h$ needed for $N=60$ was only about factor of 10.  The characteristic size of these $\beta$-functions at one-loop is typically about $0.1\%$.  So if one starts from $\zeta(t_0)=1,$ it is unlikely that the radiative corrections will give large logarithmic corrections, forcing one to take into account the running of $\zeta(t).$

The second aspect of radiative corrections concerns corrections to the effective potential which, at one-loop order, may be written as
\be
\Delta V_1=\frac{1}{64\pi^2}{\rm STr}[M(h)^4\log(M(h)^2)],
\ee
where STr denotes the ``supertrace," the sum over all the bosons minus the sum over all the fermions.  At first, this might be cause for worry, since $\lambda^2$ is so small, and there are large masses in this sum, including gauge boson masses and a top quark mass of order $h.$  However, as usual, there will be enormous cancellations between the bosons and fermions since, in the supersymmetric limit, radiative corrections to the potential vanish.  In this model, we are dealing with $F$-type susy breaking, whose characteristic order parameter is the size of $D_sW,$ because $D_d W=D_u W=0$ at $S=0.$  In fact, $D_sW=W_s=\lambda h^2/2,$ the same as in flat space.  We need an expression that expresses the result in terms of the supersymmetry breaking.  This has been given in a number of papers,
{\it e.g.,}~\cite{Srivastava:1983gb, Scholl:1984hj, Einhorn:1982pp}, and can be generically expressed as
\be
\Delta V_1=\frac{1}{64\pi^2}\sum [|f(h)|^2\log(M(h)^2)],
\ee
where $f$ is proportional to the magnitude of susy breaking and $M(h)$ are the scalar masses.  In our case,
$f=\lambda F_s=\lambda W^*_s\approx \lambda^2 h^2$ and  $M(h)\approx \lambda h.$  So up numerical factors and additive constants, we conclude
\be
\Delta V_1\approx \frac{(\lambda h)^4}{64\pi^2}\log{h^2},~{\rm so\ that}~\frac{\Delta V_1}{V_F}\approx (\frac{\lambda}{4\pi})^2 \log{h^2},
\ee
which will be completely negligible, given the small size of $\lambda$ and the limited range over which $h$ varies.

In sum, it seems quite likely that radiative corrections will not destroy this inflationary scenario.  It remains to be seen whether it is possible to marry this model  with the weak-scale NMSSM, including its other interactions and soft-breaking terms.

\section{Concluding remarks}

We presented a model of inflation where the role of the inflaton is played by the neutral part of the Higgs fields in the NMSSM supergravity model.  The model does not require non-minimal coupling to gravity, but does require an additional term in the K\"{a}hler potential that suppresses the exponential nature of the SUGRA potential and avoids the large-$\eta$ problem. The fact that the inflaton is the Higgs field makes the model more tightly constrained than more hypothetical scalar fields with properties that are usually mildly constrained by theoretical and observational considerations. As explained, the model is even more interesting and constrained in the Peccei-Quinn limit, because the current status of axion searches leave a narrow window which is in the same parameter region as that needed for inflation. 
Even though our particular model is essentially ruled out by existing CMB observations, it is interesting that models of this sort may be testable in other ways as well.  This should be kept in mind in building more sophisticated models.

Models of this type
must also be tested theoretically by determining whether the running of the parameters, such as $\tan\beta$ from the supra-Planck scale down gives sensible values for the NMSSM at the electroweak scale.  One also wants to understand the evolution of $\zeta,$ although, as the universe proceeds to temperatures well below the Planck scale, and the background field values become correspondingly smaller, its relevance becomes negligible.

Starting at $\zeta=1$ in the inflationary regime begs for a deeper explanation.  We suggested two:  a theoretical motivation from an approximate shift symmetry and a phenomenological one from the slow roll constraints, which can only be satisfied in the neighborhood of this value.  Nevertheless, one cannot help but wonder whether there is not more underlying the appearance of this shift symmetry than meets the eye.

\section{Acknowledgements}
IBD would like to thank Ramy Brustein, Shanta de-Alwis, and Misha Shifman for useful discussions, and MBE similarly thanks Tim Jones for discussions on many aspects of models such as those discussed here.  We would like to thank R. Kallosh and A. Linde for helpful correspondence.  We would also like to thank the Hebrew University of Jerusalem for its hospitality, and IBD would like to acknowledge the University of Minnesota as well, where portions of this work was carried out.  The research of IBD was supported in part by ISF grant 470/06, and that of MBE by the National Science Foundation under Grant No.\ NSF PHY05-51164.

\appendix
\section{Charged Field Fluctuations}

In this Appendix, we wish to restore the charged Higgs fields to show that fluctuations in them do not change our conclusions, in particular, that they are stable.  Since charge is conserved along the inflaton trajectory, the fluctuations in the charged fields must respect charge charge conservation.  Therefore, their lowest order nonzero terms will be quadratic in the charged fields.  As a result, they do not mix with other modes, and we can set all the other fields equal to their minimum values.  We will assume $\zeta=1$ as well.
The shift symmetry eq.~(\ref{eq:shift}) of course includes the charged fields, so the $K$-flat direction can be extended to include the charged fields.  Just as we found that $H_u^0=-H_d^{0\dagger}$ is the flat direction, one finds that $H_u^+ = H_d^{-\dagger}$ is the $K$-flat direction for the charged fields.  Indeed,
\begin{equation}
K= |H_u^+ - H_d^{-\dagger}|^2.
\end{equation}

The exact formula for the $D$-term can be written as
\begin{equation}
V_D= \frac{g_1^2+g_2^2}{8}({H_d^\dagger}H_d-{H_u^\dagger}H_u)^2 + \frac{g_2^2}{2}|{H_d^\dagger}H_u|^2.
\end{equation}
Inserting the values of the neutral fields in the K-flat direction, we find
\begin{equation}
V_D = \frac{g_1^2+g_2^2}{8}\big(|H_u^+|^2-|H_d^-|^2\big)^2 +
\frac{g_2^2 h^2}{4}|H_u^+ - H_d^{-\dagger}|^2.
\end{equation}
Thus, even though not shift invariant, it happens that $V_D=0$ also for fluctuations in the K-flat direction.

For the $F$-term, we note that, for $S=0,$ the superpotential $W$ as well as $W_u$ and $W_d$ all vanish, and $W_s=\lambda H_d H_u.$  Inserting the values of the neutral Higgs, then $W_s= \lambda[h^2/2+ H_u^+ H_d^-].$
Thus, for other fields equal to their minimum values, the F-term is simply
\begin{equation}
V_F=e^K|W_s|^2=e^{|H_u^+ - H_d^{-\dagger}|^2} \lambda^2\Big|\frac{h^2}{2}+ H_u^+ H_d^- \Big|^2.
\end{equation}

To determine local stability, we only need the expressions above to second order in $H_u^+, H_d^-.$   Note that the first term in $V_D$ is of fourth order and can be neglected, so we only need to retain the second term in $V_D.$  We need to expand $V_F$ to second order.  To simplify notation a little, we define
$H_u^+ \equiv h_+$ and $H_d^- \equiv h_-.$  (Note that, except in the flat direction, $h_-^\dagger \ne h_+$.)
Then we find
\begin{equation}
V_F=  \frac{\lambda^2 h^2}{4} \big[ h^2+ h^2 |h_+ - h_-^{\dagger} |^2 +4\Re(h_- h_+) \Big]
\end{equation}
Writing $4\Re(h_- h_+)=|h_+ + h_-^{\dagger} |^2-|h_+ - h_-^{\dagger} |^2,$ and adding back the contribution of $V_D,$ we get
\begin{equation}
V=  \frac{\lambda^2 h^2}{4}\Big[ h^2+ h^2 |h_+ - h_-^{\dagger} |^2 + |h_+ + h_-^{\dagger} |^2\Big] + (g_2^2-\lambda^2)\frac{h^2}{4}|h_+ - h_-^{\dagger} |^2.
\end{equation}
This shows that, provided $g_2>\lambda$, the charged fluctuations increase the energy, and the fluctuations along the K-flat direction and the fluctuations orthogonal to that direction decouple.
Letting $\sqrt{2}u\equiv h_+ + h_-^{\dagger}, \sqrt{2}v\equiv h_+ - h_-^{\dagger},$ we can express this result as
\begin{equation}
V=  \frac{\lambda^2 h^4}{4}+\Big[\lambda^2 h^2 +(g_2^2-\lambda^2) \Big]  \frac{h^2|v|^2}{2}  +   \frac{\lambda^2 h^2|u|^2 }{2}.
\end{equation}
Now we can read off the masses of the two charged modes
\begin{equation}
m_u^2=\frac{\lambda^2 h^2}{2}, \quad m_v^2=\frac{h^2}{2}\Big[\lambda^2 h^2 +(g_2^2-\lambda^2) \Big] .
\end{equation}
Although stable, the mass of the charged fluctuation along the flat direction ($m_u$) is of the same order as the masses $m_i$ of the scalar fluctuations  in the singlet $S.$  This  too may have to be increased in order to make them greater than the Hubble scale $H_V.$


\end{document}